\documentclass[conference,10pt]{IEEEtran}

\usepackage{algorithmic}
\usepackage[ruled, vlined]{algorithm2e}
\usepackage{header}
\usepackage{bm}
\usepackage{amsfonts}
\usepackage{comment}
\usepackage{graphicx}

\newcommand{\e}{\mathrm{e}}

\DeclareMathOperator{\deff}{def}

\begin{document}

\title{Instantaneous Frequency Estimation in Noisy Multicomponent Signals with Interfering Modes Based on Prony Method and Spline Approximation}

\author{B. Dubois-Bonnaire, S. Meignen, and K. Polisano
\thanks{The authors are with the Jean Kuntzmann Laboratory, University Grenoble Alpes and CNRS 5225, Grenoble 38401, France  (emails: basile.dubois-bonnaire@univ-grenoble-alpes.fr, sylvain.meignen@univ-grenoble-alpes.fr, kevin.polisano@univ-grenoble-alpes.fr).}}

\maketitle

\begin{abstract}
In this paper, we propose a novel estimator of the instantaneous frequencies (IFs) of the modes making up multicomponent signals (MCSs). 
We are particularly interested in dealing with noisy MCSs containing close modes 
in the time-frequency plane. 
Though it is possible to adapt Prony approach to estimate IFs in such situations, 
interference between the modes generates oscillations in the obtained estimations. 
After having investigated the nature of these oscillations, we propose an algorithm to remove these in IFs estimation, 
based on spline approximation.  
Numerical applications in various situations illustrate the benefit of mixing 
Prony technique with spline approximation for IF estimation in 
noisy MCSs containing close modes.    
\end{abstract}

\begin{IEEEkeywords}
Time-frequency, AM/FM multicomponent signal, interference, finite rate of innovation, Prony method.
\end{IEEEkeywords}

\newcommand{\cmt}[1]{{\color{red} #1}}

\IEEEpeerreviewmaketitle
\section{Introduction}
\IEEEPARstart{N}{on}-stationary signals such as audio signals (music, speech, bird songs) \cite{gribonval2003harmonic}, electrocardiogram \cite{Herry2017} and thoracic and abdominal movement signals \cite{Lin2016} can be approximated as a superimposition of amplitude and frequency-modulated (AM/FM) modes, called \emph{multicomponent signal} (MCS), and 
defined as 
\begin{equation}
\label{def:MCS}
f(t) = \sum_{p=1}^P f_p(t), \textrm{ with } f_p(t) = A_p(t)\e^{2i\pi \phi_p(t)},
\end{equation}
where the \emph{instantaneous amplitudes} (IAs) $A_p(t)$ and the \emph{instantaneous frequencies} (IFs) $\phi_p'(t)$ are supposed to be positive. To capture frequency variations over time is essential when dealing with MCSs \cite{Flandrin1998}, for which the \emph{short-time Fourier transform} (STFT)
\begin{eqnarray}
	\label{def:STFT}
	V_f^h(t, \eta) = \int_\R f(x) h(x-t) \e^{i 2\pi \eta (x-t)} \d x,
\end{eqnarray}
with $h$ a real window, is commonly used.
The spectrogram, the squared absolute value of the STFT $S_f^h(t,\eta):=\left|V_f^h(t, \eta)\right|^2$, is often used for visualization purpose.

The IFs of the modes are traditionally estimated on the spectrogram by considering local maxima along the frequency axis of that TF representation \cite{delprat1997global}, assuming the modes are well separated in the TF plane.
When two modes get too close, IF estimation using ridge computation becomes challenging 
and even impossible when some specific TF patterns, called \emph{time-frequency bubbles} (TFBs), appear \cite{meignenOneTwoRidges2022}. 

In such instances, it is possible to use Prony method \cite{deprony1795essai} to estimate the IFs of the modes by considering the spectrogram at each time instant, 
and then by using the so-called annihilating filter technique applied to some entries derived from the spectrogram \cite{blu2008sparse}.  The quality of estimation is however strongly dependent on how close the modes are in the TF plane and also  deteriorates when some noise is added to the signal.  
Inspired by \cite{blu2008sparse}, to improve IF estimation in a noisy environment, a strategy is to  denoise the entries used in the computation of the annihilating filter using Cadzow denoising \cite{cadzow1988signal}, and then apply the former or compute the IF estimates  
using a so-called \emph{total least squares approach} (TLSA) \cite{cadzow1994total}. 

However, when two modes interfere, the associated IF estimations oscillate around the true IFs, as a result of the interference pattern. As we will see, in the case of noise-free pure tones, such IF estimates intersect the true IFs at recurrent time instants corresponding to moments where the interference module cancels out.
We will see that, once detected, one can leverage these instants to obtain much better IF estimates using spline approximation. We will then investigate how the proposed approach extends to a more general context where the signal is not made of pure tones and when noise is present.

In the following section, we briefly recall how to use Prony technique for IF estimation on the spectrogram, and in what way this approach is limited when the modes interfere. We then detail a novel approach for IF estimation based on spline approximation and show its relevance in that context. 

\section{IF Estimation Based on the Prony Method and Spline Approximation}
\label{sec:Prony}  
\subsection{The Noiseless Case}
To start with, to have an idea of the interference in the TF plane, let us consider 
the signal $f$ made of two pure harmonics, i.e. $f(t)= A \e^{i 2 \pi \omega_1 t}+ \e^{i 2 \pi \omega_2 t}$. Computing its STFT with the window $h_\sigma(t) = \e^{-\pi \frac{t^2}{\sigma^2}}$, one obtains the following spectrogram
\begin{eqnarray}
\label{def:interf}	
   \begin{aligned}
  S_{f}^{h_\sigma}(t, \eta)= \sigma^{2}\Big[\overbrace{A^{2}\e^{-2\pi \sigma^{2}(\eta-\omega_{1})^{2}} + \e^{-2\pi \sigma^{2}(\eta-\omega_{2})^{2}}}^{\text{Modes part}}\label{eq:TF:2harmModePart}\\
	+ \underbrace{2A\e^{-\pi \sigma^{2}\big((\eta-\omega_{1})^{2} + (\eta-\omega_{2})^{2}\big)} \cos(2\pi(\omega_{2}-\omega_{1})t)}_{\text{Interference part}}\label{eq:TF:2harmInterfPart}\Big].
    \end{aligned}
\end{eqnarray}

When the modes are such that $|\omega_2-\omega_1|$ is large, the interference may be neglected. 
In the case of an MCS made of $P$ pure tones with constant amplitude, 
neglecting the interference in the spectrogram leads to the following approximation 
\begin{eqnarray}
\label{def:spec}	
  S_{f}^{h_\sigma}(t, \eta) \approx \sum_{p=1}^{P} a_p 
        g_\sigma (\eta -\eta_p),
\end{eqnarray}
with $g_\sigma (x) = \e^{-2\pi \sigma^2x^2}$, 
where $a_p$ approximates the squared amplitude of the $p^{th}$ mode, 
and $\eta_p$ approximates $\omega_p$. 
Assuming the amplitude and frequency of the modes vary with time, 
these can be estimated based on the Prony method \cite{deprony1795essai,rahman1987total,legros2022time}.
To start with, one computes $s_{n,k} \approx S_f^{h_\sigma}(\frac{n}{F_s}, \frac{k}{K} F_s) $ corresponding to
\begin{eqnarray} \label{eq:coeff}
    \begin{aligned}
        s_{n,k} = &\sum\limits_{p=1}^{P} a_{p,n} g\left(\frac{k}{K}F_s - \eta_{p,n}\right)\\
        =&\sum\limits_{p=1}^{P} a_{p,n} \sum\limits_{m \in \mathbb{Z}} c_m(g_{F_s}) \e^{ i 2\pi \frac{m \left(\frac{k}{K}F_s-\eta_{p,n}\right)}{F_s}}\\
        \approx & \sum\limits_{m \in {\mathbb Z}} c_m(g) 
        \underbrace{
        \sum\limits_{p=1}^{P} a_{p,n} \e^{ -i 2\pi \frac{m\eta_{p,n}}{F_s} }}_{l_{n,m}} \e^{ i 2\pi\frac{mk}{K}},
   \end{aligned}
\end{eqnarray} 
in which $g := g_\sigma$ (we drop the $\sigma$ for the sake of simplicity) and $n$ is added in 
$a_{p,n}$ and $\eta_{p,n}$ to account for their possible variations in time. In Eq. \eqref{eq:coeff}, $c_m (g_{F_s})$ is the $m^{th}$ Fourier coefficient of the restriction of 
$g$ to $[-F_s/2,F_s/2]$; since $g(\pm F_s/2)$ is very small, these can be 
approximated by $c_m (g) := \frac{1}{F_s} \hat{g} (\frac{m}{F_s})$.
We approximate the infinite sum in Eq.~\eqref{eq:coeff} by
$ 
s_{n,k} \approx \sum\limits_{m=-M_{0}}^{M_{0}} 
        c_m(g)  l_{n,m}  \e^{ i 2\pi \frac{m k}{K}}$,
which rewrites for a fixed $n$ as $  
\boldsymbol{l}_{n} = \mathbf{D}_{g}^{-1} \mathbf{V}^{-1} \boldsymbol{s}_{n}$,
where $\boldsymbol{s}_{n}=(s_{n,k})_k$,   $\mathbf{V}^{-1}$ is the left inverse of $\mathbf{V}$ and  $\mathbf{D}_{g}$ is a diagonal matrix gathering 
the Fourier coefficients $c_m(g)$ for $m = -M_{0},\dots,M_{0}$. 
Once $\boldsymbol{l}_{n}$ is computed, the Prony method is used to retrieve $\boldsymbol{\eta}_p = (\eta_{p,n})_n$:
let $\boldsymbol{h}$ be a filter of size $P+1$ such that for all $j$, $(\boldsymbol{l}_{n} \ast {\boldsymbol{h}})_j = 0$ and remark that
\begin{eqnarray} 
\label{eq:convolutionVandermonde}
    \begin{aligned}
       (\boldsymbol{l}_{n} \ast \boldsymbol{h})_j  
        &=\sum_{p=1}^{P} a_{p,n} \e^{-i 2\pi j \frac{\eta_{p,n}}{F_s}} {H\left(\e^{-i 2\pi \frac{\eta_{p,n}}{F_s}} \right)},\\ 
    \end{aligned}
\end{eqnarray}
with $H(z)$ the $\mathcal{Z}$-transform of $\boldsymbol{h}$. The expression in \eqref{eq:convolutionVandermonde} 
is null if and only if $\e^{-i 2\pi \frac{\eta_{p,n}}{F_s}}$ is a root of $H$. 
As it is preferable to consider indices $m$ with small magnitude in $l_{n,m}$ (see \cite{dubois2023instantaneous}), one writes \eqref{eq:convolutionVandermonde} for $j=1,\dots,P$, obtaining the following Yule-Walker system
\begin{equation} \label{eq:Yule-walker_system}
   \mathbf{A} \boldsymbol{h} :=
    \!\begin{pmatrix}
        \!l_{n,0} & \!\cdots & \! l_{n,-P+1} \\
         \!\vdots & \!\ddots &  \!\vdots \\
        \!l_{n,P-1} & \!\cdots & \!l_{n,0}
    \end{pmatrix}
    \!\begin{pmatrix}
        \!h_1 \\
        \!\vdots \\
        \!h_P
    \end{pmatrix}
    \!= \!-\!
    \!\begin{pmatrix}
        \!l_{n,1} \\
        \!\vdots \\
        \!l_{n,P}
    \end{pmatrix},
\end{equation}
%
which has a unique solution. 

\begin{figure}[!htb]
\begin{center}
\begin{tabular}{c c}
   \includegraphics[height = 4 cm,width=0.23 \textwidth]{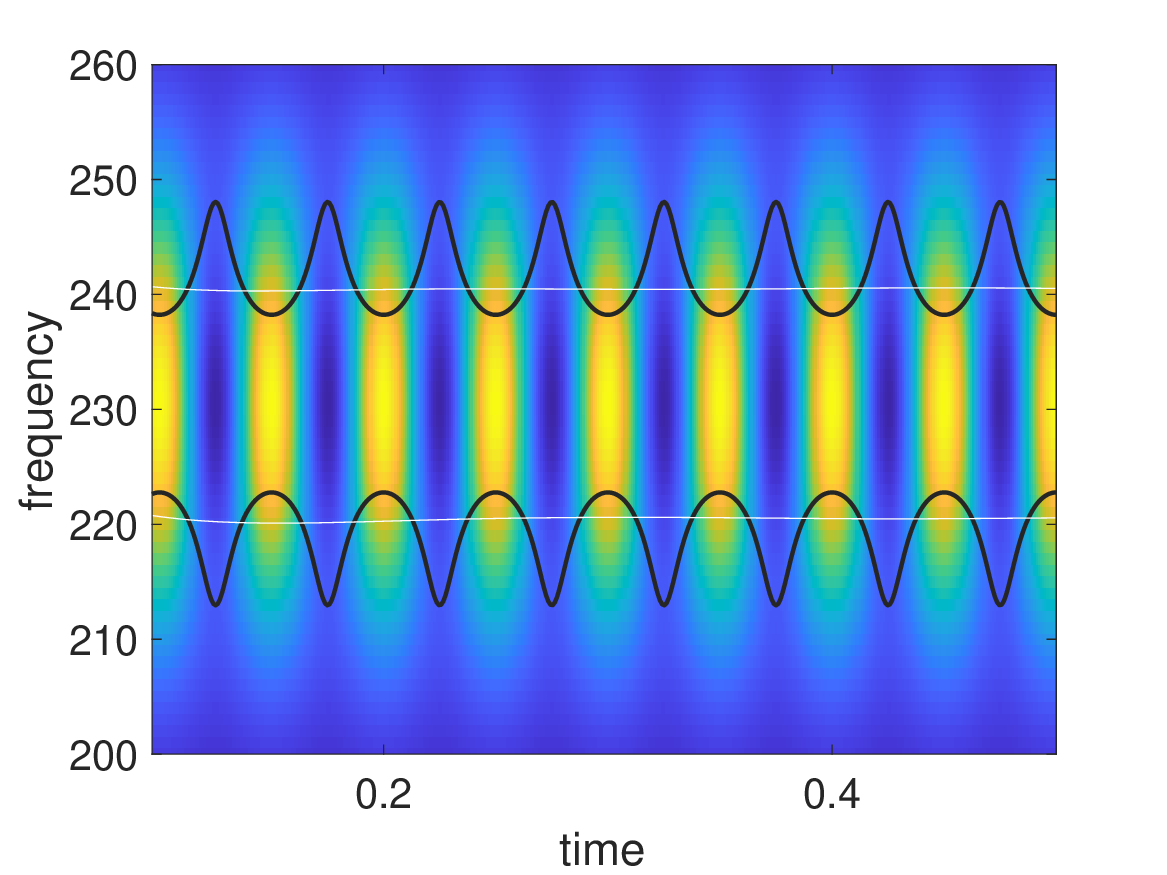} & 
   \includegraphics[height = 4 cm,width=0.23 \textwidth]{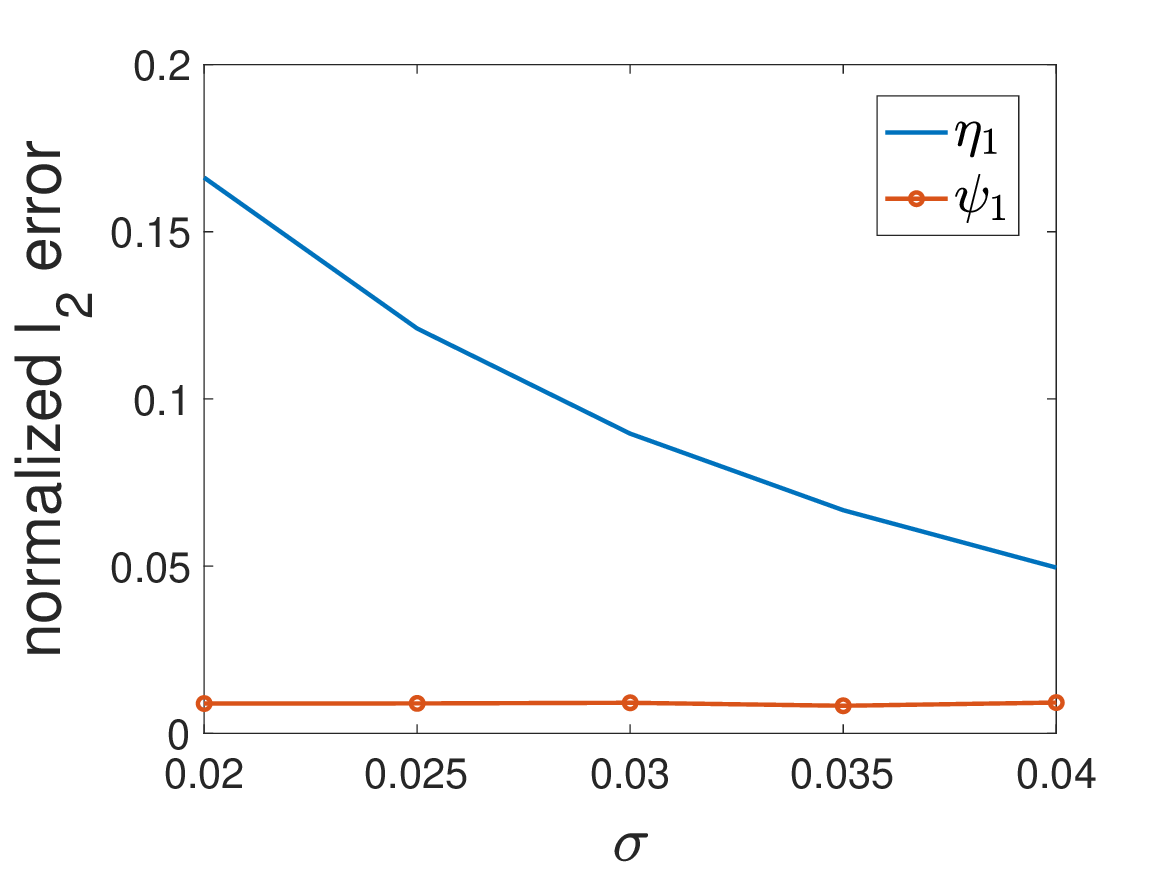}\\
   (a)&(b)
\end{tabular}		
\end{center}
\caption{(a) Spectrogram of two pure tones (same amplitude $\omega_1= 220.5$, $\omega_2 = 240.5$, $\sigma = 0.02$), and IFs estimates computed with Prony method as explained in Sec. \ref{sec:Prony} (black curves) and estimations 
$\psi_1$ and $\psi_2$ computed with \eqref{def:B_spline} (in white, $r = 1-10^{-4}$);
(b) normalized $l_2$ error associated with the estimation of 
$\omega_1$ using either $\boldsymbol{\eta}_1$ or $\boldsymbol{\psi}_1$ defined in Eq. \eqref{def:B_spline}.}
\label{Fig0}
\end{figure}

One of the main limitation of such an approach is that the interference between the modes 
create strong oscillations in the estimations $\boldsymbol{\eta}_{p}$, as illustrated in  Fig. \ref{Fig0} (a) and (b).  
To analyze these oscillations in the absence of noise, we first remark that when considering two pure tones as in  Eq.~\eqref{def:interf}, with $A=1$, at times $m_q :=\frac{q+1/2}{\omega_2-\omega_1}$, $q \in \mathbb Z$ 
and frequency $\frac{\omega_1+\omega_2}{2}$, the spectrogram is equal to zero. At these time instants, the Prony method computes 
two IFs estimates that are shifted towards higher (resp. lower) frequencies for the highest (resp. lowest) frequency mode. 
Such shifts are maximal at these time instants.   
Conversely, at time instants $M_q:=\frac{q}{\omega_2-\omega_1}$, $q \in \mathbb Z$, the spectrogram passes through a maximum along the time axis, and Prony method computes two IFs estimates that are shifted 
towards lower (resp. higher) frequencies for the highest (resp.lowest) frequency mode. Again, such shifts are maximal at these times instants. This is illustrated in Fig. \ref{Fig0} (a). 
In the middle of $[m_q,M_q]$ for any $q\in \mathbb Z$, namely at times $\frac{q+1/4}{\omega_2-\omega_1}$, the interference is not present in the spectrogram, since 
the cosine vanishes in Eq. \eqref{def:interf}, leading to an exact 
estimation of the IFs with the Prony method. Our goal is to 
estimate these time instants, and then explain how to use them  
to improve IF estimation. It is important to remark that this description of the estimation bias due to interference can be generalized to the case of $P$ modes, provided a mode interferes strongly with only one other mode at a time.   

To compute some estimates of the time instants where interference is not present 
in the spectrogram for the $p^{th}$ mode we assume, without loss of generality, 
that the signal $f$ is defined over $[0,1]$ and then discretized by a factor 
of $\frac{1}{N}$. Then, the time indices $n \in \{0,\dots,N-1\}$ 
associated with $M_q$ (resp. $m_q$), i.e. $n = \lfloor M_q N \rceil$ (resp. $n = \lfloor m_q N \rceil$), for some $q$ in $\mathbb{Z}$, and where $\lfloor X \rceil$ denotes the closest integer to $X$, are computed as the set of time indices $n$, denoted by  ${\cal I}_{\max}$ (resp. ${\cal I}_{\min}$), 
associated with local maxima (resp. minima) of the estimate of ${\boldsymbol \eta}_{p}$ given by the Prony method. Then, we define the sequence  
$\cal I := {\cal I}_{\min} \bigcup {\cal I}_{\max}$ ranked in increasing order and, finally,  
\begin{eqnarray}
\label{def:set_approx}
{\cal I}_{if} = \left\{ \left\lfloor \frac{{\cal I}(n)+{\cal I}(n+1)}{2} \right\rceil, n, \ n+1  \in \deff ({\cal I}) \right\},
\end{eqnarray}
with $\deff ({\cal I})$ the set of definition of ${\cal I}$. 
The subscript $if$ stands for "interference-free", since this set of points 
is an estimate of the time instants, in the case of two interfering pure tones, where the interference vanishes in the spectrogram.  

It is worth noting here that ${\cal I}_{if}$ makes sense only in the case of 
interfering modes, and for the IF estimation based on spline approximation 
we are going to consider, we will need extra points of interest, in the case the modes are not interfering. For that purpose, we numerically notice that the noise always generates oscillations in IFs estimation with Prony technique, and those are associated with inflection points corresponding to a sequence of time indices 
${\cal I}_{inf}$, the subscript \emph{inf} standing for inflection. 
We thus propose the following strategy to select the points of interest to build our new IF estimate. We split the interval $[0,1]$ into $Q$ 
intervals of equal length, corresponding to $I_q, \ q=0,\dots,Q-1$, 
discrete intervals of time indices and then define:
\begin{eqnarray}
\label{def:final_set}
\begin{aligned}
{\cal I}_{fin} := 
\bigcup\limits_q \left \{\begin{array}{l} 
I_q \cap {\cal I}_{if} \textrm{ if } 
I_q \cap {\cal I}_{if} \neq \emptyset \\
I_q \cap {\cal I}_{inf} \textrm{ if }  I_q \cap {\cal I}_{if} = \emptyset, 
I_q \cap {\cal I}_{inf} \neq \emptyset\\
m(I_q) \textrm{ otherwise},
\end{array}
\right .
\end{aligned}
\end{eqnarray}
with $m(I_q)$ the middle of $I_q$. 
In \eqref{def:final_set}, the first set of points of interest is considered when interference is present, the second is used where the mode is noisy but 
without interference, and the last one is useful in none of the above situations.   
We, finally, compute a new IF estimate for the $p^{th}$ mode by considering cubic spline fitting as follows:
\begin{eqnarray}
\label{def:B_spline}
\begin{aligned}
{\boldsymbol \psi}_p := \\
\mathop{\textrm{argmin}}\limits_\varphi \  \sum_{n \in {\cal I}_{fin}} 
\left|\eta_{p,n} - \varphi \left(\frac{n}{N}\right)\right|^2 + r \int_0^1 |\varphi ^{(2)}(t)|^2 \mathrm{d}t,
\end{aligned}
\end{eqnarray}
with $r \in [0, 1]$.
To illustrate the benefits of using this spline approximation for IF estimation, in the absence of noise, we display in Fig \ref{Fig0}. (b), the normalized $l_2$ error, associated with the lowest frequency mode of the two pure tone signal corresponding to the spectrogram of Fig. \ref{Fig0} (a), and when the window length parameter $\sigma$ varies.  
This error is defined by:
\begin{eqnarray}
\label{eq:def_error}
E (\boldsymbol{x},\omega_1) = \frac{1}{N} \sqrt{\sum_{n=1}^N (x_n -\omega_1)^2},
\end{eqnarray}
$x_n$ being either equal to $\eta_{1,n}$ or $\boldsymbol{\psi}_1(\frac{n}{N})$.  
The spline approximation \eqref{def:B_spline} enables to compensate for IF estimation errors with Prony method resulting from mode mixing. Note that, in this example, since 
the modes are always interfering ${\cal I}_{fin} = {\cal I}_{if}$.             
The main problem with the proposed spline approximation is that it is built on the IF estimation given by Prony technique which is irrelevant in noisy situations, and 
we investigate how to cope with this in the following section.   

\subsection{The Noisy Case}
\label{sec: Prony_noise}
We assume, in the following, that the signal $f$ is contaminated by a Gaussian noise $\varepsilon$,  with zero mean and unknown variance $\sigma_\varepsilon^2$, to obtain 
$\tilde f = f + \varepsilon$. 
In the sequel, we denote by $\tilde s_{n,m}$, the spectrogram of the noisy signal $\tilde f$, and then $\tilde l_{n,m}$ the coefficients obtained from $\tilde s_{n,m}$ following the same approach as in the noiseless case.
To denoise $\tilde l_{n,m}$, a common strategy, 
known as Cadzow denoising \cite{blu2008sparse,cadzow1988signal}, 
consists of considering the following square Toeplitz matrix (with  $P\leq T \leq M_0$):
\begin{equation} \label{eq:Cadzow}
    \tilde{\mathbf{B}}_T  \! = \! 
    \begin{pmatrix}
        \!\tilde l_{n,0} & \!\cdots & \! \tilde l_{n,-T}\\
          \!\vdots & \!\ddots &  \!\vdots \\
        \!\tilde l_{n,T}& \!\cdots & \! \tilde l_{n,0}
    \end{pmatrix},
\end{equation}
whose rank is $P$ in the absence of noise and $T+1$ otherwise. 
Then one computes the  \emph{singular value decomposition} (SVD) of  
$\tilde {\mathbf B}_{T} = {\tilde{\mathbf U}} {\tilde {\mathbf \Sigma}} {\tilde{\mathbf W}}^*$, in which $^*$ denotes the Hermitian transpose. The SVs in $\tilde {\mathbf \Sigma}$ being ranked in decreasing order with respect to their amplitude, one 
defines a new matrix  
${\tilde{\mathbf B}}_{T,P} = {\tilde{\mathbf U}_P} {\tilde {\mathbf \Sigma}}_P {\tilde{\mathbf W}_P}^*$, with  ${\tilde{\mathbf U}_P}$ (resp ${\tilde{\mathbf W}_P^*}$) corresponding to the first $P$ columns (resp. rows) of $\tilde{\mathbf U}$ (resp $\tilde{\mathbf W}^*$). Such a matrix is of rank $P$ but no longer Toeplitz. 
To retrieve this structure for the matrix, one replaces the coefficients on each diagonal of $\tilde{\mathbf B}_{T,P}$ 
by the average of the coefficients on this diagonal, to obtain the matrix $\tilde{\mathbf B}_{T,P}^{[1]}$. One then iterates this procedure until the $(P+1)^{th}$ SV is smaller than the $P^{th}$ by some prerequisite factor. 

One can then solve the Yule-Walker system \eqref{eq:Yule-walker_system} using the denoised values of $ \boldsymbol{\tilde l}_n$, and finally compute the IFs estimates. This technique is referred to as \emph{cad} (for Cadzow) in the sequel.
As suggested in \cite{blu2008sparse}, an alternative technique is to replace the Yule-Walker system by a \emph{total least square approximation} (TLSA), by considering first the following rectangular matrix: 
\begin{equation} \label{eq:TLSA}
    \tilde {\mathbf{A}}_T  \!= \! 
    \begin{pmatrix}
        \!\tilde l_{n,-T+P} & \!\cdots & \! \tilde l_{n,-T}\\
             \!\vdots & \!\ddots &  \!\vdots \\
        \! \tilde l_{n,T} & \!\cdots & \! \tilde l_{n,T-P}
    \end{pmatrix},
\end{equation}
and then searching for a minimizer of $\|\tilde {\mathbf{A}}_T\boldsymbol{h}\|^{2}$, constrained by $\|\boldsymbol{h}\|^{2}=1$. 
This is performed by computing the SVD of $\tilde{\mathbf{A}}_T$ \cite{cadzow1988signal}, and then by setting $\boldsymbol{h}$ to be the eigenvector associated with the smallest singular value. In the simulations that follow, we will denote this technique by \emph{cad-tlsa} (Cadzow denoising followed by total least square). 

As mentioned in Sec. \ref{sec:Prony}, interference between modes creates oscillations in the estimated IFs. Though it is essential 
to denoise $ \boldsymbol{\tilde l}$, 
this denoising procedure does not remove these oscillations. 
To get rid of them, we follow the same framework as in the noiseless case, but we first remove potential outliers in IF estimation obtained using \emph{cad} or \emph{cad-tlsa}, 
by not allowing jumps in IFs estimation. 
To bridge the gaps between the time instants where outliers are detected, we use \emph{piecewise cubic monotone Hermite interpolation} (pchip) \cite{fritsch1980monotone}. Then, we define the new set 
${\cal I}_{fin}$ associated with this interpolation signal, 
and the spline approximation following the same approach as in Sec. \ref{sec:Prony}. As this procedure can be applied either after \emph{cad} or \emph{cad-tlsa}, this generates  
two different techniques denoted by \emph{cad-spline} or \emph{cad-tlsa-spline}, in the following.  
The whole procedure is summarized in Algorithm 1
\begin{minipage}{\columnwidth}
\vspace{1ex}
\rule{\columnwidth}{.3ex}\\*
{\textbf{Algorithm 1:} IFs Estimation using Prony method and spline approximation}\\*
\rule[.5ex]{\columnwidth}{.3ex}\\*
\textbf{Input:} -- $\mathbf{\tilde s}$ noisy spectrogram of MCS with $P$ (known) modes  \\
\vspace{-3 ex}
\begin{algorithmic}[1]
\STATE Denoise $\boldsymbol{\tilde l}$ using Cadzow denoising. 
\STATE Estimate IFs using annihilating filter or TLSA.
\STATE Compute pchip interpolation \cite{fritsch1980monotone} of  these estimations after outliers removal. 
\STATE Improve IFs estimation using \eqref{def:B_spline} on  interpolation signal. 
\end{algorithmic}
\vspace{-1.4ex}
\vspace{1ex}
\textbf{Output:} IF estimates.\\
\rule[1ex]{\columnwidth}{.3ex}
\end{minipage}
\begin{figure}[!htb]
\begin{center}
\begin{tabular}{c c}
   \includegraphics[height = 4 cm,width=0.24 \textwidth]{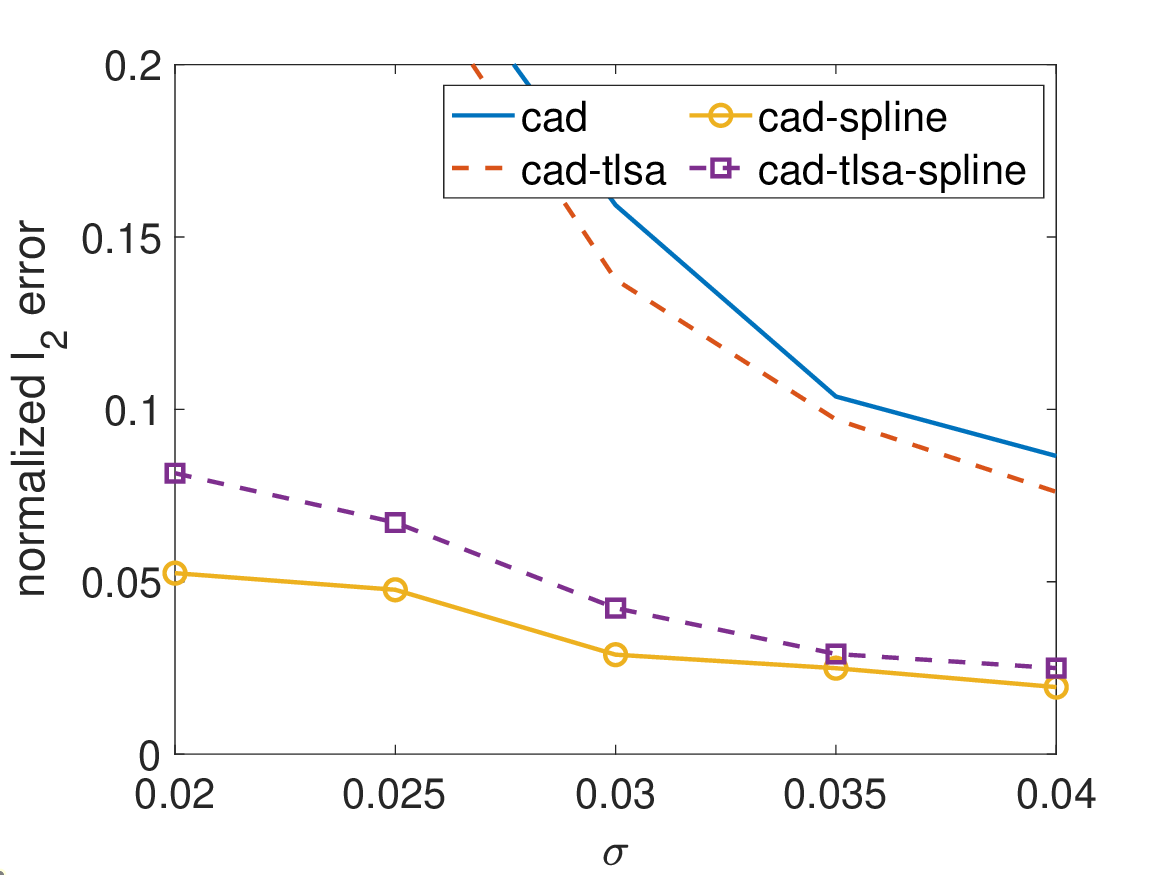} & \includegraphics[height = 4 cm, width= 0.24\textwidth]{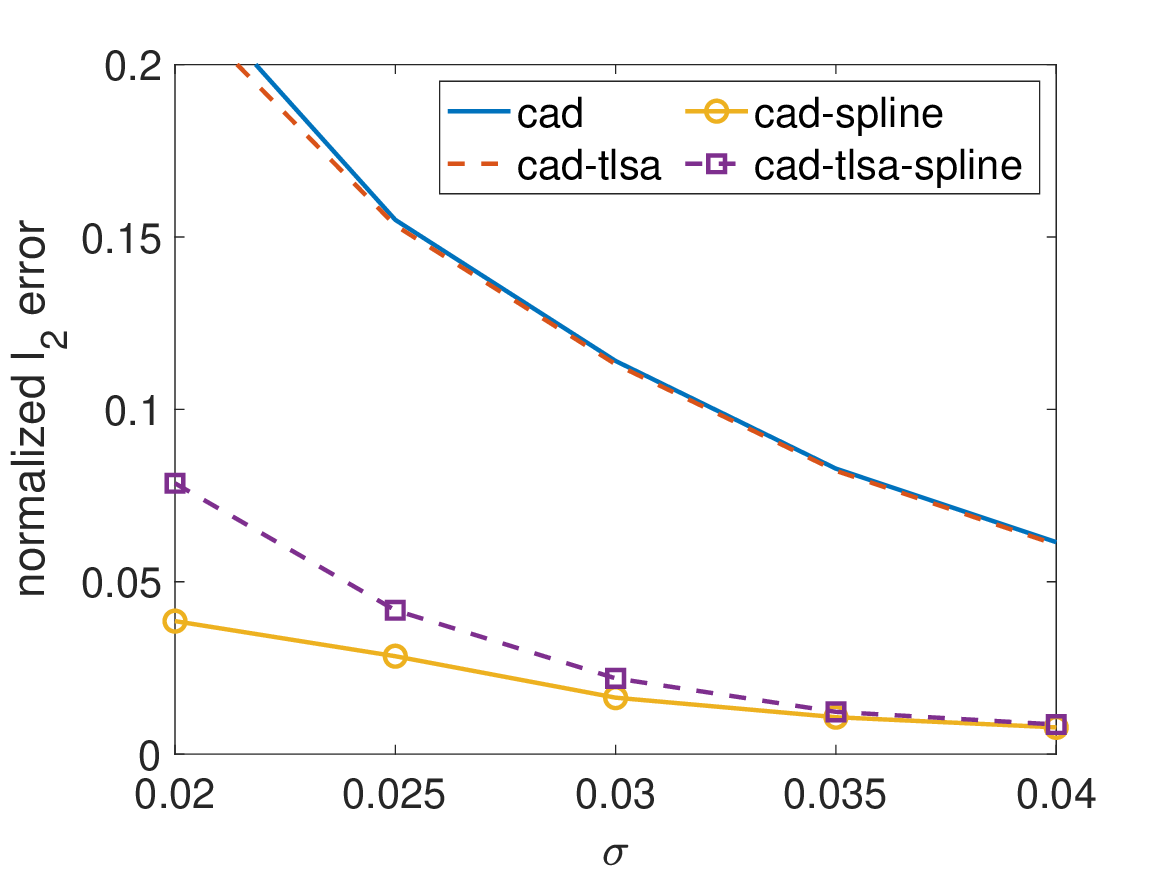}\\
   (a)&(b)
\end{tabular}		
\end{center}
\caption{(a) Normalized $l_2$ error of the mode $f_1$ of  the 
signal of Fig. \ref{Fig1} (a), for varying $\sigma$ using different techniques for IF estimation: cad, cad-tlsa, cad-spline, 
and cad-tlsa-spline (input SNR is 5 dB, the results are averaged over 10 noise realizations); 
(b) same as (a) but for an input SNR of 10 dB.} 
\label{Fig1}
\end{figure}

To illustrate the behavior of our IF estimation technique, we consider the same signal as in Fig. \ref{Fig0}, when both $\sigma$ and the input SNR vary. At medium SNR, 
the results of Fig. \ref{Fig1} (b) suggest that,  \emph{cad} and \emph{cad-tlsa} lead to the same quality of estimation as in the noiseless case by  
comparing with Fig. \ref{Fig0} (b). Thus, both techniques are efficient for denoising, but not for oscillations removal. 
On the contrary, the estimation results are considerably improved 
by considering \emph{cad-spline} or  \emph{cad-tlsa-spline}, 
with a slightly better performance when the spline approximation is carried out on the estimation given by \emph{cad} technique. 
Note that the benefits of using spline approximation is even more significant when the noise level increases, as shown in Fig. \ref{Fig1} (a). 

\section{Results}
\label{sec:results}

\subsection{The Case of a Two parallel Linear Chirps}
\label{sec:two_chirp}
\begin{figure}[!htb]
\begin{center}
\begin{tabular}{c c}
   \includegraphics[height = 4 cm,width=0.24 \textwidth]{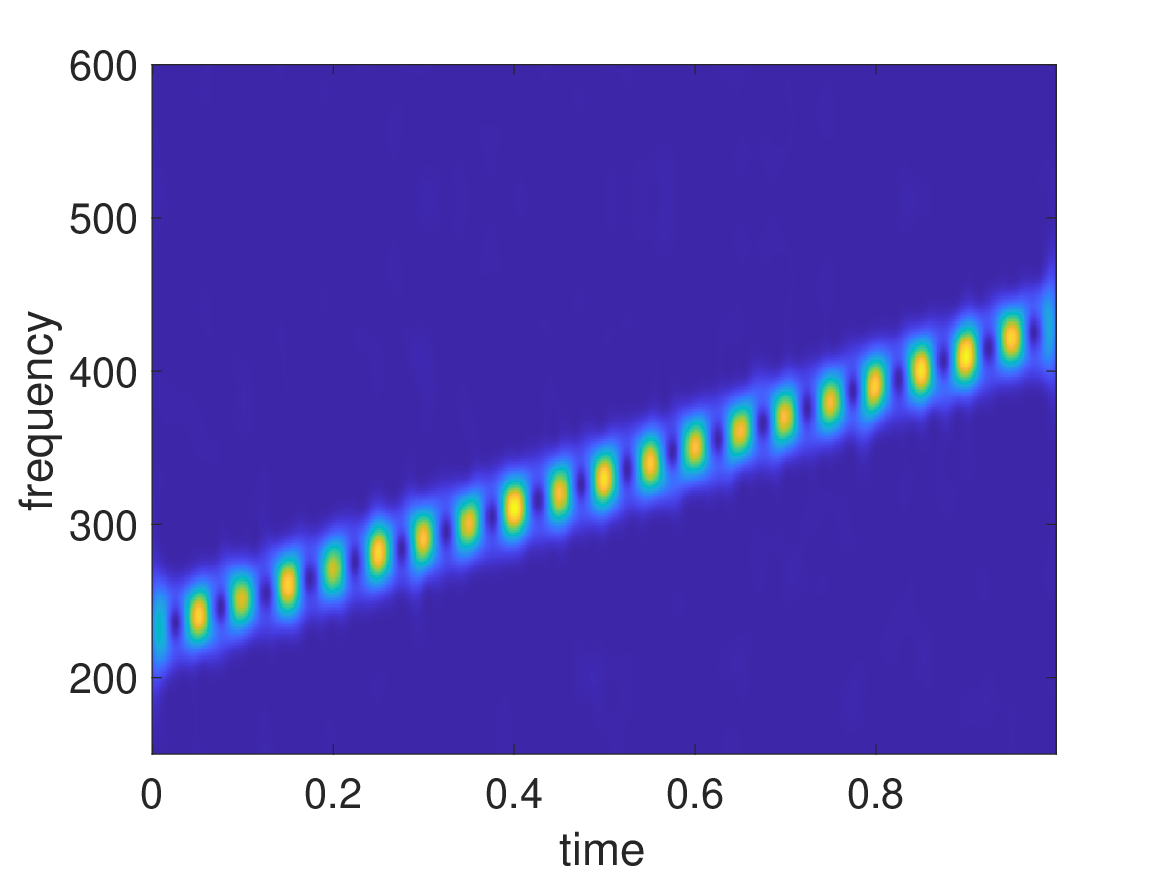} & \includegraphics[height = 4 cm, width= 0.24\textwidth]{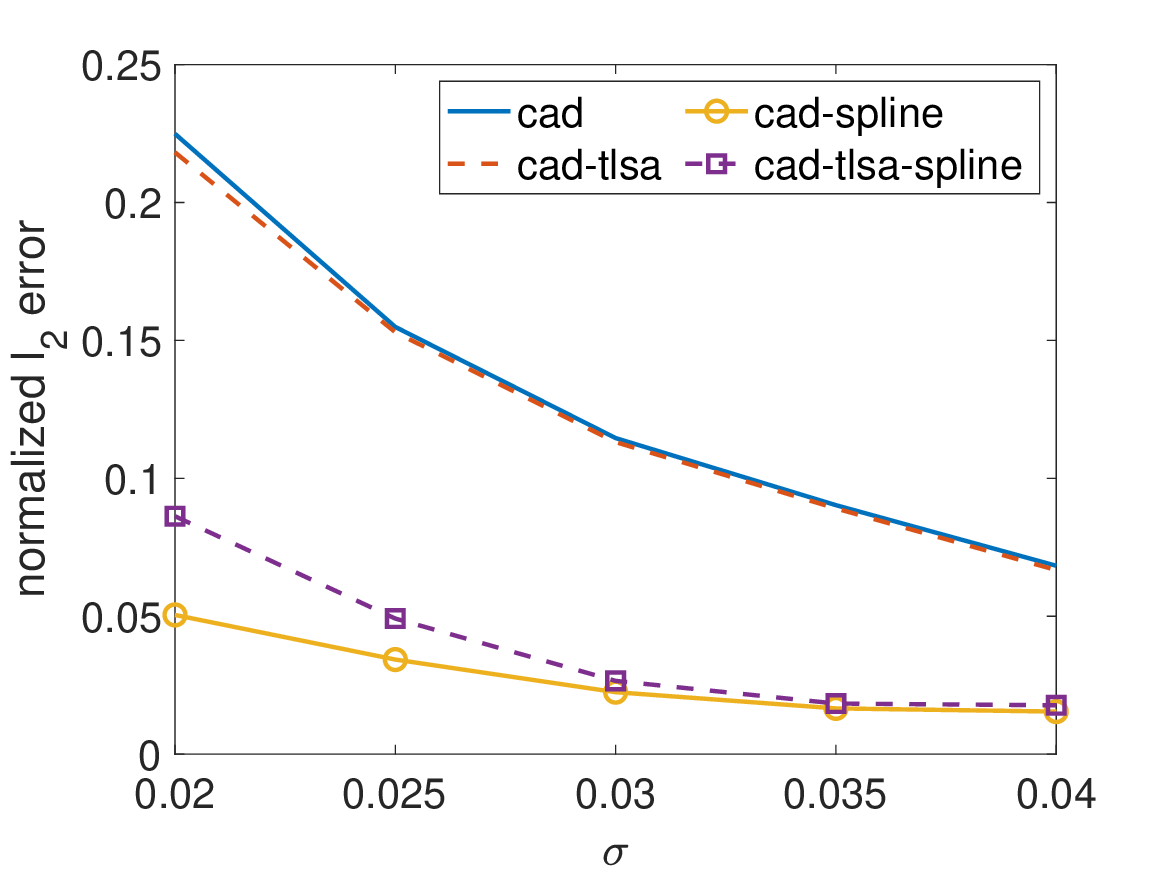}\\
   (a)&(b)
\end{tabular}		
\end{center}
\caption{(a) Spectrogram of two parallel linear chirp (SNR=10  dB, $\sigma=0.025$); 
(b) normalized $l_2$ error for mode $f_1$ of the signal in (a), associated with \emph{cad}, \emph{cad-tlsa}, \emph{cad-spline}, and  \emph{cad-tlsa-spline} (results averaged over 10 noise realizations).} 
\label{Fig2}
\end{figure}

Here we investigate the quality of IF estimation for a signal made of two interfering parallel linear chirps, when either \emph{cad}, \emph{cad-tlsa} or 
the spline approximation introduced in Algorithm 1 are used. An illustration of the spectrogram of such a signal is given in Fig. \ref{Fig2} (a), and the normalized $l_2$ error associated with the different IF estimators is given in Fig. \ref{Fig2} (b) (input SNR = 10 dB). Comparing that figure with the case of pure tones with 
the same noise level, i.e. Fig. \ref{Fig1} (b), we notice that the algorithm behaves similarly in both cases. Note that changing the noise level would lead to the same conclusion.
In such a case, the modes are always interfering and thus 
${\cal I}_{fin} = {\cal I}_{if}$.

\subsection{More General Signals}
\label{sec:two_general_chirps}
\begin{figure}[!htb]
\begin{center}
\begin{tabular}{c c}
   \includegraphics[height = 4 cm,width=0.24 \textwidth]{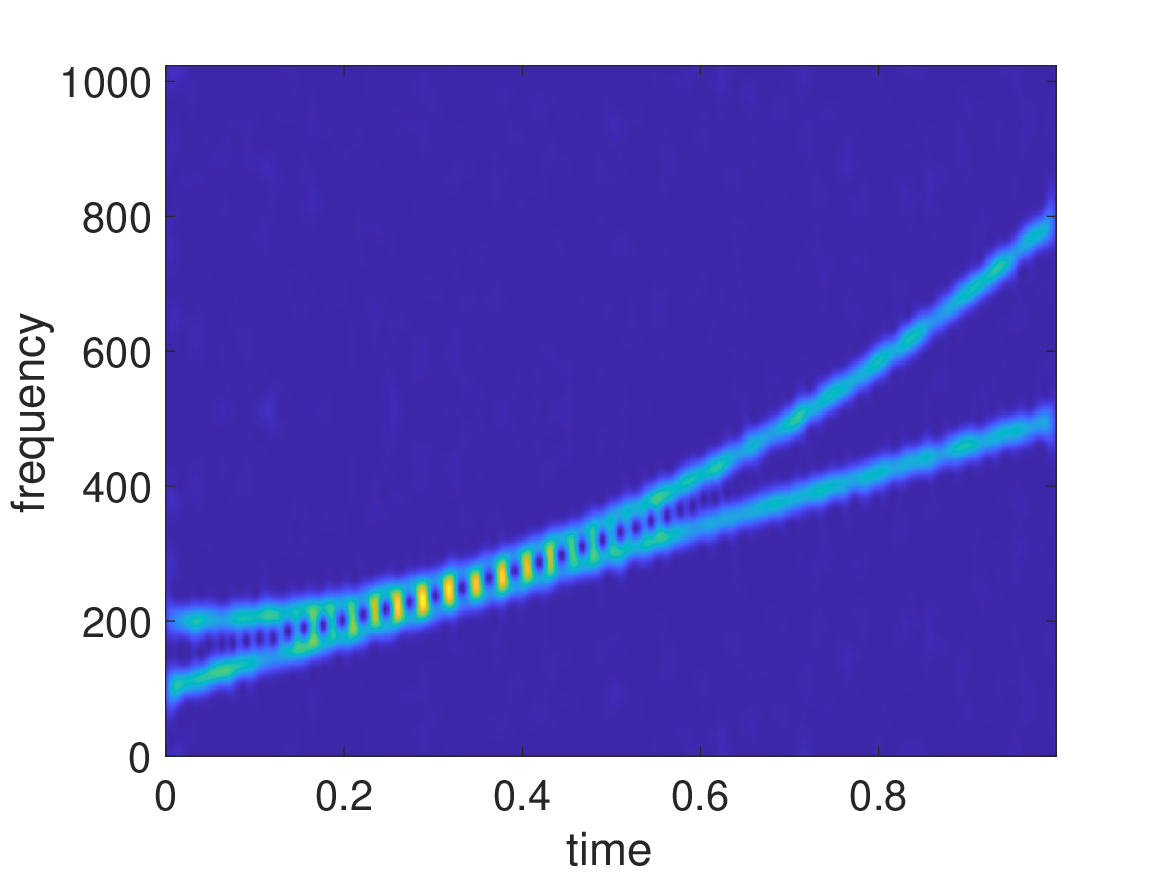} & \includegraphics[height = 4 cm, width= 0.24\textwidth]{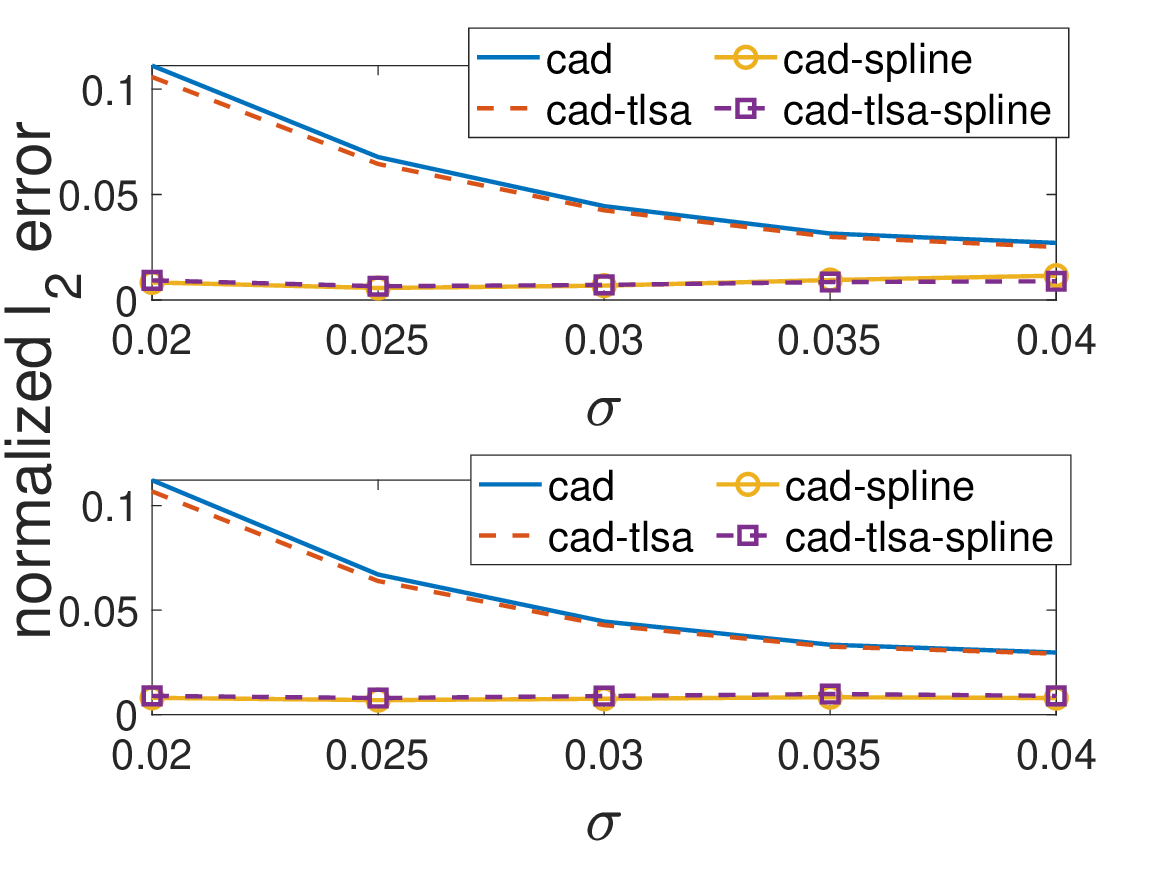}\\
   (a)&(b)
\end{tabular}		
\end{center}
\caption{(a) spectrogram of two chirps (SNR =10 dB, $\sigma=0.025$); 
  (b) normalized $l_2$ error for modes of the signal in (a) (top $f_2$, bottom $f_1$) associated with \emph{cad}, \emph{cad-tlsa}, \emph{cad-spline}, and  \emph{cad-tlsa-spline} (results averaged over 10 noise realizations).} 
\label{Fig3}
\end{figure}

In this section, we investigate the behavior of our algorithm when 
the modes are not interfering for the whole time span, 
as illustrated in Fig \ref{Fig3} (a), so that the points of interest used to build the spline approximation are not only related to the interference, but also to the inflection points of the estimation,  
when the modes are not interfering. The normalized $l_2$ error associated with IF estimation for the two modes making up 
the signal is depicted in Fig. \ref{Fig3} (b) (input SNR = 10 dB) showing the relevance of the novel estimation technique we propose in more complex situations.
\section{Conclusion}
In this paper, our goal was to propose a new technique based on spline approximation and Prony method to estimate the instantaneous frequencies of the modes of a multicomponent signal when 
the former are interfering and in noisy situations. 
The relevance of the proposed approach over classical IF 
estimators also based on Prony technique is demonstrated over 
a series of meaningful examples.    
\bibliographystyle{IEEEtran}
\bibliography{Ridges.bib}

\begin{thebibliography}{10}
\providecommand{\url}[1]{#1}
\csname url@samestyle\endcsname
\providecommand{\newblock}{\relax}
\providecommand{\bibinfo}[2]{#2}
\providecommand{\BIBentrySTDinterwordspacing}{\spaceskip=0pt\relax}
\providecommand{\BIBentryALTinterwordstretchfactor}{4}
\providecommand{\BIBentryALTinterwordspacing}{\spaceskip=\fontdimen2\font plus
\BIBentryALTinterwordstretchfactor\fontdimen3\font minus
  \fontdimen4\font\relax}
\providecommand{\BIBforeignlanguage}[2]{{%
\expandafter\ifx\csname l@#1\endcsname\relax
\typeout{** WARNING: IEEEtran.bst: No hyphenation pattern has been}%
\typeout{** loaded for the language `#1'. Using the pattern for}%
\typeout{** the default language instead.}%
\else
\language=\csname l@#1\endcsname
\fi
#2}}
\providecommand{\BIBdecl}{\relax}
\BIBdecl

\bibitem{gribonval2003harmonic}
R.~Gribonval and E.~Bacry, ``Harmonic decomposition of audio signals with
  matching pursuit,'' \emph{IEEE Transactions on Signal Processing}, vol.~51,
  no.~1, pp. 101--111, 2003.

\bibitem{Herry2017}
C.~L. Herry, M.~Frasch, A.~J. Seely, and H.-T. Wu, ``Heart beat classification
  from single-lead {ECG} using the synchrosqueezing transform,''
  \emph{Physiological Measurement}, vol.~38, no.~2, pp. 171--187, 2017.

\bibitem{Lin2016}
Y.-Y. Lin, H.-T. Wu, C.-A. Hsu, P.-C. Huang, Y.-H. Huang, and Y.-L. Lo, ``Sleep
  apnea detection based on thoracic and abdominal movement signals of wearable
  piezoelectric bands,'' \emph{IEEE journal of biomedical and health
  informatics}, vol.~21, no.~6, pp. 1533--1545, 2017.

\bibitem{Flandrin1998}
P.~Flandrin, \emph{Time-frequency/time-scale analysis}.\hskip 1em plus 0.5em
  minus 0.4em\relax Academic Press, 1998, vol.~10.

\bibitem{delprat1997global}
N.~Delprat, ``Global frequency modulation laws extraction from the gabor
  transform of a signal: A first study of the interacting components case,''
  \emph{IEEE transactions on speech and audio processing}, vol.~5, no.~1, pp.
  64--71, 1997.

\bibitem{meignenOneTwoRidges2022}
\BIBentryALTinterwordspacing
S.~Meignen, N.~Laurent, and T.~Oberlin, ``One or {{Two Ridges}}? {{An Exact
  Mode Separation Condition}} for the {{Gabor Transform}},'' vol.~29, pp.
  2507--2511. [Online]. Available:
  \url{https://ieeexplore.ieee.org/document/9970392/}
\BIBentrySTDinterwordspacing

\bibitem{deprony1795essai}
G.~R. de~Prony, ``Essai experimental et analytique: sur les lois de la
  dilatabilite des fluides elastique et sur celles de la force expansive de la
  vapeur de l'eau et de la vapeur de l'alkool, a differentes temperatures,''
  \emph{Journal Polytechnique ou Bulletin du Travail fait a l'Ecole Centrale
  des Travaux Publics}, 1795.

\bibitem{blu2008sparse}
T.~Blu, P.-L. Dragotti, M.~Vetterli, P.~Marziliano, and L.~Coulot, ``Sparse
  sampling of signal innovations,'' \emph{IEEE Signal Processing Magazine},
  vol.~25, no.~2, pp. 31--40, 2008.

\bibitem{cadzow1988signal}
J.~A. Cadzow, ``Signal enhancement-a composite property mapping algorithm,''
  \emph{IEEE Transactions on Acoustics, Speech, and Signal Processing},
  vol.~36, no.~1, pp. 49--62, 1988.

\bibitem{cadzow1994total}
------, ``Total least squares, matrix enhancement, and signal processing,''
  \emph{Digital Signal Processing}, vol.~4, no.~1, pp. 21--39, 1994.

\bibitem{rahman1987total}
M.~Rahman and K.-B. Yu, ``Total least squares approach for frequency estimation
  using linear prediction,'' \emph{IEEE Transactions on Acoustics, Speech, and
  Signal Processing}, vol.~35, no.~10, pp. 1440--1454, 1987.

\bibitem{legros2022time}
Q.~Legros and D.~Fourer, ``Time-frequency ridge estimation of multi-component
  signals using sparse modeling of signal innovation,'' \emph{arXiv preprint
  arXiv:2212.11343}, 2022.

\bibitem{dubois2023instantaneous}
B.~Dubois-Bonnaire, S.~Meignen, and K.~Polisano, ``Instantaneous frequency
  estimation in multicomponent signals in case of interference based on the
  prony method,'' \emph{arXiv preprint arXiv:2312.14500}, 2023.

\bibitem{fritsch1980monotone}
F.~N. Fritsch and R.~E. Carlson, ``Monotone piecewise cubic interpolation,''
  \emph{SIAM Journal on Numerical Analysis}, vol.~17, no.~2, pp. 238--246,
  1980.

\end{thebibliography}
\end{document}